\documentclass[aoas]{imsart}

\RequirePackage{amsthm,amsmath,amsfonts,amssymb}
\RequirePackage[authoryear]{natbib}
\usepackage[flushleft]{threeparttable}
\usepackage{array,booktabs,makecell}
\RequirePackage{graphicx}

\startlocaldefs

\endlocaldefs

\begin{document}

\begin{frontmatter}
\title{Precision Dose-finding Cancer Clinical Trials in the Setting of Broadened Eligibility}
\runtitle{Precision Dose-Finding Design}

\begin{aug}
\author[A]{\fnms{Rebecca B.}~\snm{Silva}\ead[label=e1]{rs4025@cumc.columbia.edu}},
\author[A]{\fnms{Bin}~\snm{Cheng}\ead[label=e2]{bc2159@cumc.columbia.edu}}, 
\author[B]{\fnms{Richard D.}~\snm{Carvajal}\ead[label=e3]{rdc2150@cumc.columbia.edu}}
\and
\author[A]{\fnms{Shing M.}~\snm{Lee}\ead[label=e4]{sml2114@cumc.columbia.edu}}

\address[A]{Department of Biostatistics, Columbia University Mailman School of Public Health, \\ 
\printead{e1,e2,e4}}
\address[B]{Herbert Irving Comprehensive Cancer Center, Columbia University\printead[presep={,\ }]{e3}}
\end{aug}

\begin{abstract} 
Broadening eligibility criteria in cancer trials has been advocated to represent the true patient population more accurately. While the advantages are clear in terms of generalizability and recruitment, novel dose-finding designs are needed to ensure patient safety. These designs should be able to recommend precise doses for subpopulations if such subpopulations with different toxicity profiles exist. While dose-finding designs accounting for patient heterogeneity have been proposed, all existing methods assume the source of heterogeneity is known and thus pre-specify the subpopulations or only allow inclusion of a few patient characteristics. We propose a precision dose-finding design to address the setting of unknown patient heterogeneity in phase I cancer clinical trials where eligibility is expanded, and multiple eligibility criteria could potentially lead to different optimal doses for patient subgroups. The design offers a two-in-one approach to dose-finding by simultaneously selecting patient criteria that differentiate the maximum tolerated dose (MTD) and recommending the subpopulation-specific MTD if needed, using marginal models to sequentially incorporate patient covariates. Our simulation study compares the proposed design to the naive approach of assuming patient homogeneity and our design recommends multiple doses when heterogeneity exists and a single dose when no heterogeneity exists. The proposed dose-finding design addresses the challenges of broadening eligibility criteria in cancer trials and the desire for a more precise dose in the context of early phase clinical trials.  
\end{abstract}


\end{frontmatter}

\section{Introduction}
The goal of a dose-finding phase I cancer clinical trial is to select a safe dose of a new therapy for subsequent phase II and III trials. The recommended dose is often known as the maximum tolerated dose (MTD), assuming toxicity and efficacy of a treatment increase with dosage. The majority of phase I trials recommend one MTD, assuming all eligible patients have similar tolerability to the new treatment. Over the past decade, the expansion of eligibility criteria for cancer trials which were initially designed for cytotoxic chemotherapies has been advocated. Eligibility criteria have been found to be too restrictive and not reflective of the actual patient population that could benefit from a therapy [\cite{fehrenbacherRandomizedClinicalTrial2009, gerberImpactPriorCancer2014, malikEligibilityCriteriaPhase2019, liuEvaluatingEligibilityCriteria2021, harveyImpactBroadeningTrial2021}], thus limiting the generalizability of study results and significantly slowing down patient accrual [\cite{malikEligibilityCriteriaPhase2019}].  One study examining the ineligibility rates of 326 patients consecutively diagnosed with non-small cell lung cancer (NSCLC) found that about 80\% of patients are ineligible based on traditional eligibility criteria [\cite{fehrenbacherRandomizedClinicalTrial2009}].

Several authors have found that the inclusion of patients with brain metastases, HIV, or prior cancers, does not affect trial outcomes and better reflects the actual patient population [\cite{georgeReducingPatientEligibility1996, kimModernizingEligibilityCriteria2015, kimBroadeningEligibilityCriteria2017, beaverReevaluatingEligibilityCriteria2017a, malikEligibilityCriteriaPhase2019}]. In 2017, the American Society of Clinical Oncology (ASCO), Friends of Cancer Research, and the US Food and Drug Administration published new recommendations for exclusion criteria that should be reconsidered including brain metastases, HIV and AIDS, organ dysfunction, prior or concurrent cancers, and a minimum age for enrollment [\cite{kimBroadeningEligibilityCriteria2017, fda}]. Moreover, eligibility criteria such as concomitant medications, prior therapies, performance status, liver or renal dysfunction, and laboratory values such as levels of bilirubin, platelets, hemoglobin and alkaline phosphatase, continue to be questioned and evaluated by ASCO and Friends of Cancer Research [\cite{harveyImpactBroadeningTrial2021, liuEvaluatingEligibilityCriteria2021, magnusonModernizingClinicalTrial2021, osarogiagbonModernizingClinicalTrial2021}].

While the broadening of eligibility criteria has clear advantages, it also has to be done cautiously to ensure patient safety.  Specifically, dose-finding methods should account for the potential of heterogeneous populations where patient subpopulations could have different tolerances and take a precision medicine approach to dose estimation by recommending a subpopulation-specific MTD if needed.  Methods have been proposed in the phase I setting to account for a pre-specified patient covariate or subpopulations. All these methods either assume that the covariates of interest or subpopulations with different toxicity profiles are known beforehand, or adaptively combine subgroups based on similar risk without the identification of covariates. Therefore, these methods cannot be applied to address the challenges in the setting of broadened eligibility criteria where it is not known beforehand whether different MTDs are needed.  Extensions of the Continual Reassessment Method (CRM) [\cite{[piantidosi1996]}] and the Escalation with Overdose Control allow for the addition of a patient-specific covariate to the dose-toxicity model [\cite{babbPatientSpecificDosing2001, chengIndividualizedPatientDosing2004}]. Moreover, there are methods that allow for the pre-specification of subpopulations including the two-sample CRM [\cite*{oquigleyTwosampleContinualReassessment1999}], the two-sample CRM for ordered groups [\cite{oquigleyContinualReassessmentMethod2003}], and isotonic designs for multiple ordered risk groups [\cite{yuan_chappell2004}] and partially ordered risk groups [\cite{conawayIsotonicDesignsPhase2017}]. More recently, \citet*{moritaSimulationStudyMethods2017} proposed a hierarchical Bayesian approach that uses subgroup-specific intercepts and \citet{chappleSubgroupspecificDoseFinding2018} proposed Spike and Slab priors on subgroup parameters to account for multiple groups. 

While precision medicine has been generally advocated for later stages in drug development and with larger data sets, in our early stage of drug development, it would be ideal to have a method that can screen for the expanded eligibility criteria, assess whether it is necessary to account for them, and be able to recommend more than one MTD if needed. \citet{beaverReevaluatingEligibilityCriteria2017a} propose that “a logical approach to defining eligibility could allow for detection of safety signals in early clinical trials that use broad eligibility criteria”. As eligibility criteria in oncology trials expand, having a method that can learn and assess for signals of patient heterogeneity in toxicity will be key to ensuring the safety of patients who have traditionally been thought to be at increased risk of adverse outcomes.

As a motivating example, we use the phase I trial evaluating intermittent dosing of a mitogen-activated protein kinase kinase (MEK) inhibitor, Selumetinib, to treat patients with uveal melanoma [\cite{khanIntermittentMEKInhibition2022}]. The study evaluated six doses of Selumetinib (100, 125, 150, 175, 200, 225 mg) and estimated the MTD using the Time-to-Event CRM (TITE-CRM) with a target toxicity level of 25\% in 28 patients. Given the rarity of uveal melanoma, accrual took over three years with three large sites. The principal investigator of the trial identified three criteria (creatinine levels, organ and marrow function, and the presence of known or suspected brain metastases) that could have been loosened to allow more patients to benefit from the trial. For example, creatinine level  captures renal function and between 12\% and 53\% percent of cancer patients have chronic kidney disease [\cite{kitchluRepresentationPatientsChronic2018}]. If eligibility criteria were to expand, the design of this phase I trial would require accounting for these three criteria, since knowledge about their association with toxicity is limited. Some of these criteria may affect the probability of toxicity of Selumetinib while others may not. For example, renal impairment has been associated with increased risk of toxicity, leading investigators to account for degree of renal dysfunction when determining dose [\cite{lameireNephrotoxicityRecentAnticancer2014, lealDoseescalatingPharmacologicalStudy2011, lorussoPharmacokineticsSafetyBortezomib2012}]. However, dose reductions have been found to not always be necessary [\cite{lealDoseescalatingPharmacologicalStudy2011}]. If the loosened criteria are not accounted for and at least one criterion is associated with increased risk of toxicity, the design could overestimate the MTD for these patients.

Our proposed design, the Precision CRM (P-CRM), addresses the case where we want to consider and learn about multiple patient characteristics, which contribute to a broader patient population. For example, if the eligibility criteria were expanded in the Selumetinib trial, we would want to consider evaluating three patient criteria since the MTD could differ based on one or more of these criteria. Our method selects patient criterion to add to the dose-toxicity model sequentially, based on their marginal correlation with toxicity after accounting for dose. With each new enrollment cohort, given the additional information on toxicity and patient characteristics, the criteria are tested for inclusion or removal. Dose assignment is based on the selected patient criteria given the current data and the final subpopulation-specific MTD is determined using the selected covariates in the final model after the total sample size is reached. This method provides a two-in-one approach to dose-finding; the method learns about the criteria and the need to differentiate the MTD and recommends a dose for each patient subpopulation based on these criteria.

The rest of the paper is outlined as follows. The precision dose-finding method is described in Section~\ref{s:methods}. In Section~\ref{s:sim}, we present a simulation study based on the redesigned Selumetinib trial and compare the performance of the P-CRM to the naive approach of not accounting for any patient heterogeneity. We also discuss sample size considerations for the design. Finally, the significance and limitations of our proposed design are discussed in Section~\ref{s:discuss}.

\section{Methods}
\label{s:methods}

Drug safety is most commonly measured by the occurrence of a severe adverse event, or toxicity, in the first cycle of treatment, known as a dose-limiting toxicity (DLT). In model-based designs, a total of $J$ doses, $d_1, \ldots, d_J$, are evaluated, and the probability of toxicity, or DLT, is estimated using a dose-toxicity model. The MTD, defined  as the dose whose toxicity is closest to $p_T$, the highest rate of toxicity that clinicians deem acceptable, or target toxicity level, is estimated sequentially. For all future notation, let $Y_i$ be the binary toxicity outcome for the $i$th patient and $X_i$ be the dose assigned to this patient. 

\subsection{Dose-toxicity Model}

Traditionally, it is assumed that the entire patient population has the same probability of toxicity at each dose, and therefore has the same MTD. Assuming a logistic model and complete patient homogeneity, $\pi(d_j,\beta_0,\beta_1) = P(Y_i = 1 | X_i=d_j)$ is modeled as
\begin{equation}
    \text{logit}\{\pi(d_j, \beta_0, \beta_1)\}= \beta_0 + \beta_1 d_j, j=1, \ldots, J. \label{eqn-base}
\end{equation}

In the late 1990s, the patient homogeneity assumption in dose-finding models started being questioned. A dose-toxicity model incorporating a patient covariate, $z$, was proposed, where $\pi(d_j, z_i, \boldsymbol{\theta}) = P(Y_i = 1 | X_i=d_j, z_i)$ is modeled as,
\begin{equation}
    \text{logit}\{\pi(d_j, z_i, \boldsymbol{\theta})\}= \beta_0 + \beta_1 d_j + \gamma z_i, \label{eqn-one-cov}
\end{equation}
where ${\boldsymbol \theta} = (\beta_0, \beta_1, \gamma)^T$ [\cite{[piantidosi1996]}]. 
Equation (\ref{eqn-one-cov}) represents an overly restrictive model; it assumes that we know beforehand which covariate contributes to the heterogeneity and that heterogeneity is completely due to $z$, both of which are strong assumptions.

In this paper, we consider a potentially large collection of patient covariates, $\boldsymbol{z} = ({z}_1, ..., {z}_M)^T$, which could be responsible for dose heterogeneity. We model $\pi(d_j, \boldsymbol{z}_i, \boldsymbol{\theta}) = P(Y_i = 1 | X_i=d_j, \boldsymbol{z}_i)$ as
\begin{equation}
\text{logit}\{\pi(d_j,\boldsymbol{z}_i, \boldsymbol{\theta})\} = \beta_0 + \beta_1 d_j + \sum_{m = 1}^M \gamma_m z_{mi} \text{,}
\label{eqn-true}
\end{equation}
where $\boldsymbol{\theta} = (\beta_0, \beta_1, \gamma_1,..., \gamma_M)^T.$ In addition, we make the following sparsity assumption
\begin{eqnarray} \sum_{m = 1}^M 1_{\gamma_m \ne 0} = K \ll M. \label{sparse} \end{eqnarray}

 In the setting of expanded eligibility criteria, Equation (\ref{eqn-true}) represents a more realistic model which is potentially high dimensional, but we assume that only a small subset of these criteria are responsible for the heterogeneity. Assumption (\ref{sparse}) is consistent with  clinicians' belief that there are only a few expanded criteria/covariates or none at all for which we need to actually differentiate the MTD.

Therefore, assuming there are $M$ total patient covariates suspected to be associated with toxicity of which a small subset actually differentiate the MTDs, our objective is two-fold. One is  to identify the subset of $K$ covariates that are responsible for the heterogeneity in MTDs, and the other is to implement a procedure that determines the subpopulation-specific MTDs. 

\subsection{Precision Dose-finding Procedure}

Let $\Gamma=\{ \gamma_i, i=1, \ldots, M\}$, and $ \Gamma_0= \{ \gamma_i: \gamma_i \ne 0\}$. By Equation (\ref{eqn-true}), $ \Gamma_0$ is an unknown small subset of $\Gamma$ that must be estimated because it is responsible for the heterogeneity in MTDs.  The estimation of $\Gamma$ imposes a large challenge. Since $K$ is  unknown, and thus must be estimated, we need to search  among all $2^M$ subsets of $\Gamma$ for $\Gamma_0$.  Furthermore, in dose-finding studies, patients are  assigned to doses in sequential cohorts, not simultaneously to all doses as done in a parallel group design. This means that information concerning parameters in model (\ref{eqn-true}) is to be accumulated gradually as more and more toxicity data become available, suggesting that it would be too ambitious to estimate $\Gamma_0$ in a single effort. 

Consistent with the sequential nature of dose finding studies, we propose to estimate $\Gamma_0$ sequentially. The proposed design, which we call the Precision CRM (P-CRM), comprises of two stages. In the first stage, patient characteristic information is collected, but dose assignment is not patient-specific. In the second stage, once enough information is obtained, the method sequentially selects patient covariates to add to the dose-toxicity model based on their marginal correlation with toxicity after accounting for dose level. With each new enrollment cohort, given the additional information on toxicity and patient characteristics, the selected covariate or covariates are tested for removal. Please note these are enrollment cohorts and not dose cohorts, and thus patients within a cohort are assigned to different dose levels depending on their patient characteristics.  Dose assignment is then based on the selected patient covariates given the current data. The final covariate-specific MTD is determined using the selected covariates in the final model after the total sample size is reached. The dose-finding stages of the P-CRM, in more detail, are as follows.

\textit{Stage I}: For the first $N_1$ patients, the regular one-sample CRM method based on working model (\ref{eqn-base}) is used [\cite{oquigleyContinualReassessmentMethod1990}]. Covariate information is collected but not used until Stage II. The skeleton can be elicited from physicians or calibrated through the approach by Lee and Cheung (2009) after specifying a target toxicity level, $p_T$, cohort size, and initial guess of MTD. The dose label, or standardized dose, $d_j$, for the $j$th dose is determined through equation  $\text{logit}(p_{0j}) = \tilde \beta_0 +\tilde \beta_1 d_j$, where $\tilde \beta_0$ and $\tilde \beta_1$ are the prior guesses of parameters $\beta_0$ and $\beta_1$ in Model (\ref{eqn-base}), respectively, and $p_{0j}$ is the skeleton, $j=1, \ldots, J$. The dose labels are updated using a new skeleton based on the posterior probabilities of toxicity from $N_1$ patients.

\textit{Stage II}: After $N_1$ patients have been assigned and dose and toxicity are observed, patient covariates are sequentially assessed for inclusion after each patient enrollment cohort.  
\begin{enumerate}
    \item[1.] In this stage, patients are enrolled in cohorts of size $L$. 
    \item[2.] Once the first cohort is enrolled, we start the selection of covariates. For each covariate $z_{m}, m = 1,...,M$, we fit the following working model
\begin{equation}
    \text{logit}\{\pi(d_j, z_{m,i}, \boldsymbol{\theta_m})\}= \beta_{0m} + \beta_{1m}d_j + \gamma_{m}z_{m,i} \label{eqn-marginal}
\end{equation}
and denote the p-value associated with covariate $z_m$ as $p_1(z_m)$. Let $m_1=\arg\min_m p_1(z_m)$. If $p_1(z_{m_1})\ge  \alpha^*_1$ for some prespecified $\alpha^*_1>0$, then no covariate is added and dose assignment is conducted as in Stage I. If $p_1(z_{m_1})< \alpha^*_1$, then covariate $z_{m_1}$ is selected, and the $(n+1)$-th cohort of patients is  assigned to the dose closest to the target toxicity level $p_T$:
    \begin{equation}
  x_{[n+1]}=      \text{argmin}_{x \in \{d_1,\ldots,d_J\}} | \pi(x;\boldsymbol{z}_{\boldsymbol{m_1}, n+1}, \boldsymbol{\hat{\theta}}_{\boldsymbol{m_1},n}) - p_T|. \label{eqn-argmin}
    \end{equation}

\item[3.] Once the data from a new enrollment cohort are collected, we proceed to select the next covariate and consider removal of the selected covariate. For each of the remaining covariates $z_{m}$, $m \ne m_1$, we fit working model (\ref{eqn-marginal}). Let $p_2(z_m)$ be the p-value associated with covariate $z_m, m\ne m_1$, and let $m_2=\arg\min_{m \ne m_1} p_2(z_m)$. Covariate $z_{m_2}$ is selected if and only if $p_2(z_{m_2})< \alpha^*_2$ for some prespecified $\alpha_2^*$. 

Similarly, given that $q$ covariates $z_{m_1}, \ldots, z_{m_q}$ are selected, to select the $(q+1)$-th covariate, the remaining $M-q$ covariates are tested one by one under Model (\ref{eqn-marginal}) where $m \notin \{m_1, \ldots, m_q\}$. Let $p_{q+1}(z_m)$ be the p-value associated with covariate $z_m$ in Model (\ref{eqn-marginal}), and let $m_{q+1}=\arg\min_{m \notin \{m_1, \ldots, m_q\}}p_{q+1}(z_m)$. Then, covariate $z_{m_{q+1}}$ is selected if and only if $p_{q+1}(z_{m_{q+1}})< \alpha_{q+1}^*$ for some  $\alpha_{q+1}^*>0$. 

\item[4.] To assess for removal of the selected covariates, supposing $q$ covariates $z_{m_1}, \ldots, z_{m_q}$ have been selected, we fit a working model 
\begin{equation}
    \text{logit}\{\pi(d_j, z_{m,i}, \boldsymbol{\theta_m})\}= \beta_{0m} + \beta_{1m}d_j + \sum_{l=1}^q \gamma_{m_l}z_{m_l,i}. \label{eqn-marginal_d}
\end{equation}

Let $m_q^*=\arg\max_{m} \{ p_q(z_{m_1}), \ldots, p_q(z_{m_q})\}$. Covariate $z_{m_q^*}$ is removed if $p_q(z_{m_q^*}) > \alpha_q^{**}.$ 

\item[5.] Given $q$ remaining covariates that have not been removed, dose assignment follows from Equation (\ref{eqn-argmin}) based on working model (\ref{eqn-marginal_d}). If after removal, no covariates remain, dose assignment is conducted as in Stage I. 

\item[6.]  Repeat Steps 3 to 5 until $N_{\max}$ patients are assigned a recommended dose in this manner. \\

\noindent Once $N_{\max}$ is reached, estimate the MTD as follows: 

If no covariates remain selected, the MTD is estimated using the one-sample CRM and a single MTD is recommended for all patients. If $k>0$ covariates $z_{m_1}, \ldots, z_{m_k}$ remain in the model, let $\boldsymbol{z}_k^*=(z_{m_1}, \ldots, z_{m_k})^T$, then the MTD is
  \begin{equation}
  \text{MTD}_{\boldsymbol{z}_k^*}= \text{argmin}_{x \in \{d_1,...,d_J\}} | \pi(x;\boldsymbol{z}_k^*, \boldsymbol{\hat{\theta}}_{\boldsymbol{m},N_{\max}}) - p_T|, \label{eqn-mtd}
    \end{equation}
    where $\pi(x;\boldsymbol{z}_k^*, \boldsymbol{\hat{\theta}}_{\boldsymbol{m},N_{\max}})$ comes from the final model
    \begin{equation}
    \text{logit}\{\pi(d_j, \boldsymbol{z}^*_{k,i}, \boldsymbol{\theta_m})\}= \beta_{0m} + \beta_{1m}d_j + \sum_{l=1}^k \gamma_{m_l}z_{m_l,i}. \label{eqn-final}
\end{equation}

\end{enumerate}

Regarding the choice of cohort size, $L$, we recommend using three to five patients depending on the total sample size and number of criteria being considered, $M$. The inclusion thresholds $\alpha_q^*$ and exclusion thresholds $\alpha^{**}_q, q=1, \ldots, M$, should depend on the number of marginal models that are fit for a new cohort. If $M$ is the total number of covariates being tested in the trial and $q$ is the number of covariates already selected in the model, we recommend using $\alpha^*_q=\alpha (M-q)/{M}$ for inclusion and $\alpha_q^{**}=\alpha/q$ for exclusion, with $\alpha = 0.20$. Our suggested choice of the inclusion threshold $\alpha^*_q$ is a decreasing function in $q$, indicating that we put more penalty for including more covariates, in consistency with the sparsity assumption (\ref{sparse}). 


\section{Numerical Studies}
\label{s:sim}

\subsection{Simulation Setting: Redesigning the Selumetinib study}
\label{s:sim1}
We performed a simulation study to examine the operating characteristics of the P-CRM dose-finding design. We redesigned the Selumetinib trial to consider expanding three patient criteria and evaluated the need for a subpopulation-specific MTD. The design was compared to the one-sample CRM to demonstrate how ignoring patient heterogeneity if it exists affects MTD recommendation. 

We tested three binary patient criteria ($M = 3$), and examined five scenarios of dose-toxicity relationships with six dose levels ($J=6$). For each scenario, we considered each criterion having a prevalence of 50\% and 25\%, $P(z_m = 1) = 0.50, 0.25$, for $m = 1,2,3$. The dose-toxicity relationships for each scenario are illustrated in Figure~\ref{scenarios-figure} and the exact probability of DLT at each dose is provided in the Appendix. None of the scenarios were generated from the dose-toxicity models specified previously.

The scenarios include cases when zero or one of the three eligibility criteria are strongly associated with toxicity, leading to one or two subgroups with different dose-toxicity relationships, respectively. In Scenarios 1 through 4, $z_2$ is associated with toxicity and the MTD for newly eligible patients satisfying the loosened criterion, $z_2 = 1$, differs from the MTD for patients who meet the originally published criterion, $z_2 = 0$. In both Scenarios 1 and 2, the MTDs of the subpopulations differ by one dose but in Scenario 1, the lower MTD is dose level one while in Scenario 2 the lower MTD is dose level two. These scenarios explore how the method performs when the MTDs exist on the boundary of the range of dose levels versus inside the range of dose levels. In Scenarios 3 and 4, the MTDs of the subpopulations differ by two and three dose levels, respectively, representing increasingly more heterogeneous subpopulations and a greater need to differentiate the MTDs. Lastly, in Scenario 5, no criteria are associated with toxicity so no heterogeneity exists and the MTD is the same for all patients.

In Stage I of the P-CRM and in the one-sample CRM, we started the CRM at dose level two and a target toxicity level, $p_T$, of $0.25$, as in the original trial. We used the logistic dose-toxicity model with a fixed intercept of 3 and assumed a normal prior of mean $0$ and variance $1.34$ on the dose covariate, $\beta$. The dose-toxicity model was calibrated to select a dose that yields between a 17\% and 33\% DLT rate [\cite{leeModelCalibrationContinual2009}].  The cohort size was three.  In the P-CRM, we set $N_1 = 15$ and after $N_1$ patients were assigned a dose using the CRM, we obtained the updated scaled dose as $d_j = \log (\frac{p^*_{0j}}{1-p^*_{0j}}) - 3; j = 1,...,J$, where $p^*_{0j}$ is the posterior probability of toxicity at dose $j$, assuming a dose effect coefficient of 1 and intercept of $3$. In Stage II, we used a fixed-intercept of $3$ in each marginal logistic regression model and a cohort size of three. Although three patients were added at a time, dose-assignment depended on their respective covariate pattern and the covariates chosen in the dose-finding algorithm. Additionally, we did not allow skipping a dose that had not yet been assigned to a patient. We ran 2,000 simulations of the Selumetinib trial with four different total sample sizes of $N_{\text{max}} = 30, 45, 60, 72$ to evaluate the impact of sample size. 


\begin{figure}[ht]
\centering
\includegraphics[width=1.0\textwidth]{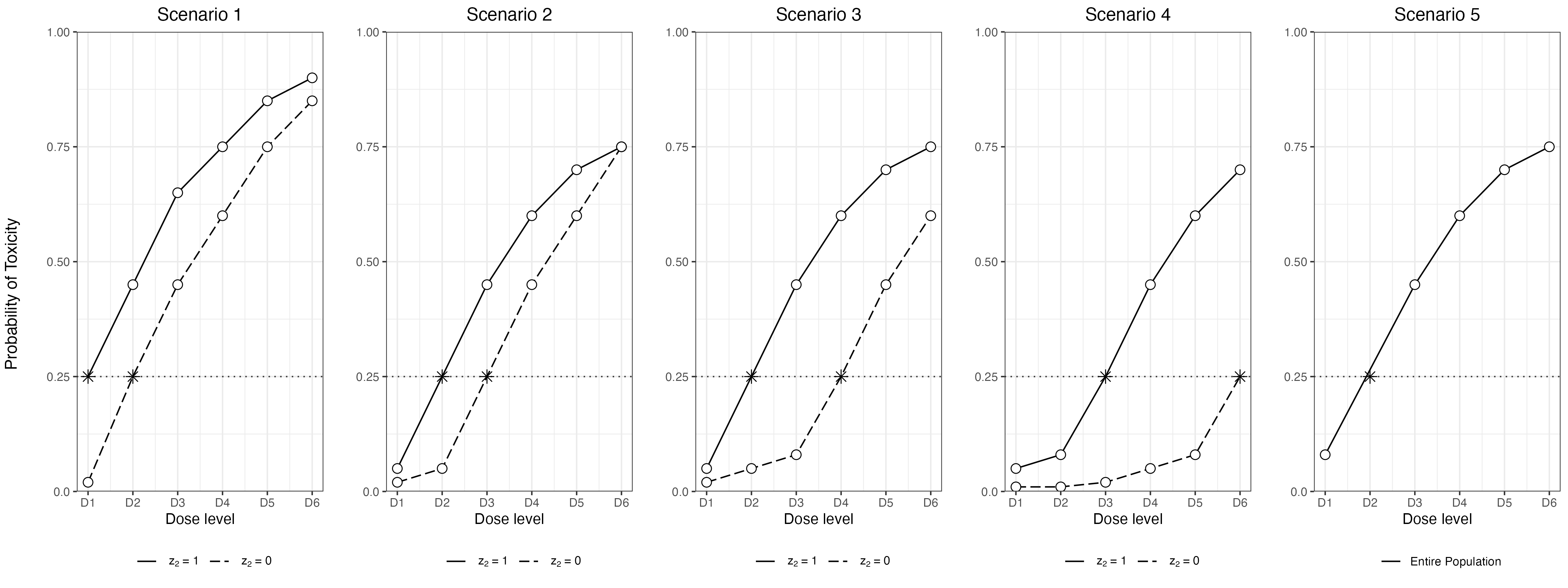}
\caption{Dose-toxicity relationships for Scenarios 1-5. Scenarios 1, 2, 3, and 4 have one criterion associated with toxicity, differentiating the MTD by one, two, or three doses levels. In Scenario 5, no criteria are associated with toxicity so there is one MTD for the population. Horizontal dotted line represents the target toxicity level of 0.25. Star-shaped point represents the covariate-specific MTD.}
\label{scenarios-figure}
\end{figure}


For each scenario and criteria prevalence, we were interested in three performance measures: the probability of correct covariate selection, the probability of correct MTD selection (PCS), and the weighted probability of MTD selection (WPS) [\cite{moritaSimulationStudyMethods2017}]. The probability of correct covariate selection was defined as the frequency in which the $k$ covariates remaining in the final model from Equation~\ref{eqn-final} were the true criteria associated with toxicity. For Scenarios 1 through 4, the correct final model includes one additional covariate besides dose, $z_2$, and for Scenario 5 it includes no additional covariates. The PCS was obtained by estimating the MTD for the observed patients based on the covariates chosen in the final model at $N_{\max} = 30, 45, 60, 72$. For each observed patient covariate pattern, we used the final model obtained from the dose-finding procedure to estimate the covariate-specific probability of toxicity at each dose and estimated the covariate-specific MTD as discussed in Section~\ref{s:methods}. If no covariates were chosen, the one-sample CRM was used to estimate the MTD. To obtain a weighted probability of selection, we used the measure defined by \citet{moritaSimulationStudyMethods2017}, which accounts for dose selection close to the MTD using a weight characterized by the distance in probability of toxicity at the selected dose and the target toxicity level. Let $r_{j,k}$ be the true probability of DLT at dose level $j$, $j = 1,\ldots,6$ for the $k$th subgroup, where subgroups are defined by the true MTD as shown in Figure 1 such that $k = 1,2$ for Scenarios 1 through 4, and $k = 1$ for Scenario 5, ordered by increasing true MTD, and $p_T$ is the target toxicity level. Then the WPS for subgroup $k$, $\text{WPS}_{k}$ is defined as
$$\text{WPS}_{k} = \sum_{j = 1}^J w_{j,k} P(d_j \text{ is selected as the MTD for subgroup $k$}),$$
where 
$$w_{j,k} = \frac{(\max_{j'=1,\ldots, J}|r_{j',k} - p_T|) - |r_{j,k} - p_T|}{(\max_{j'=1, \ldots, J}|r_{j',k} - p_T|)  - (\min_{j' = 1,\ldots, J} |r_{j',k} - p_T|)}.$$

The WPS gives no weight to the least desirable dose and relatively higher weight for doses yielding true toxicity rates closer to the target toxicity level, while the PCS gives no weight to all doses except the true MTD. The subgroup PCS and WPS from the P-CRM were compared to the subgroup PCS and WPS from the one-sample CRM. 

\subsection{Simulation Results}
\label{s:results}
\begin{figure}
\centering
\includegraphics[width=1.0\textwidth]{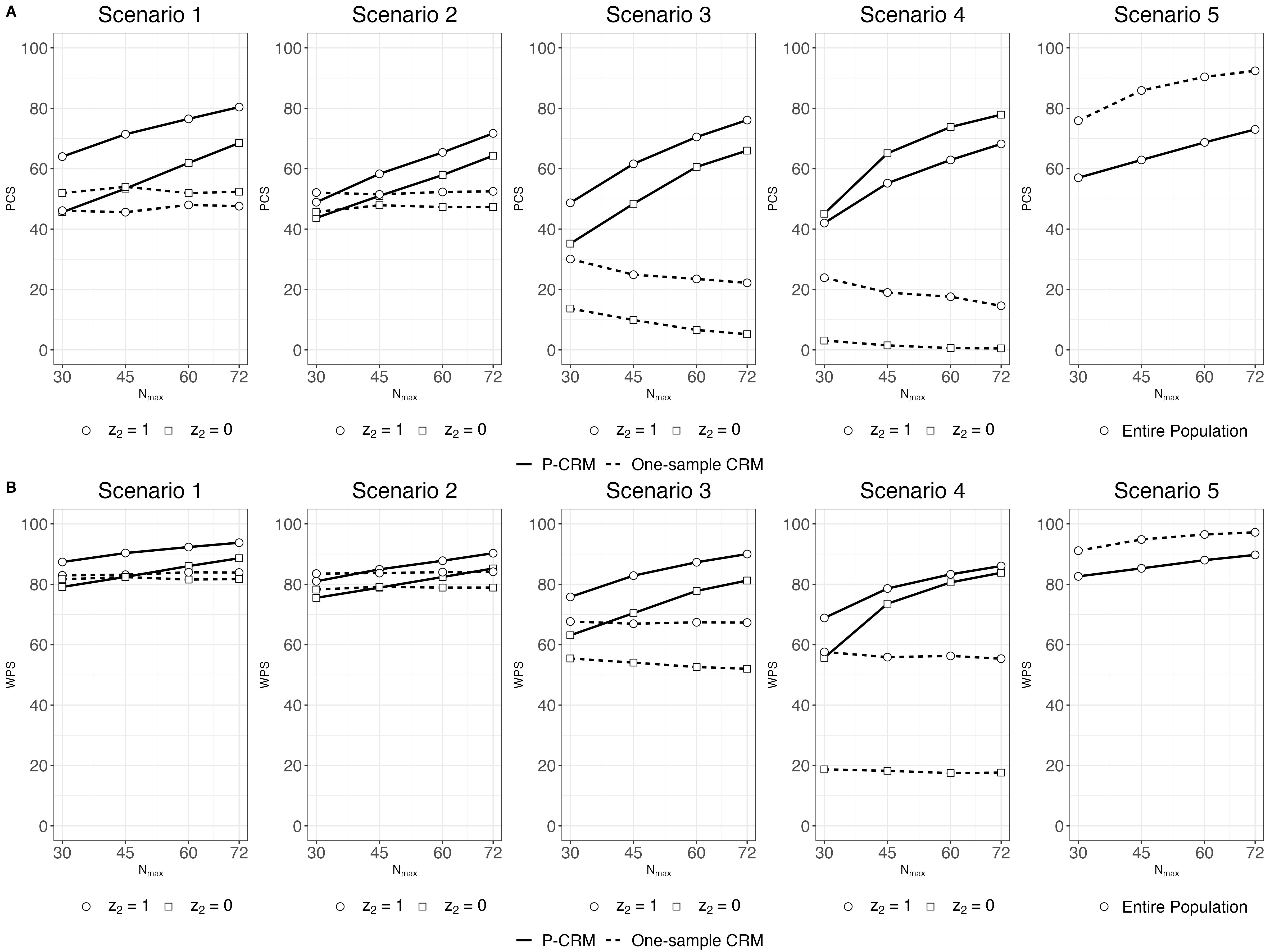}
\caption{A) Proportion of Correct Selection of MTD (PCS) and B) Weighted Probability of Selection (WPS) for each subgroup under the P-CRM and one-sample CRM across $N_{\max} = 30, 45, 60, 72$ for criteria prevalence of 50\%.}
\label{pcs_wps_fig50}
\end{figure}

\begin{figure}
\centering
\includegraphics[width=1\textwidth]{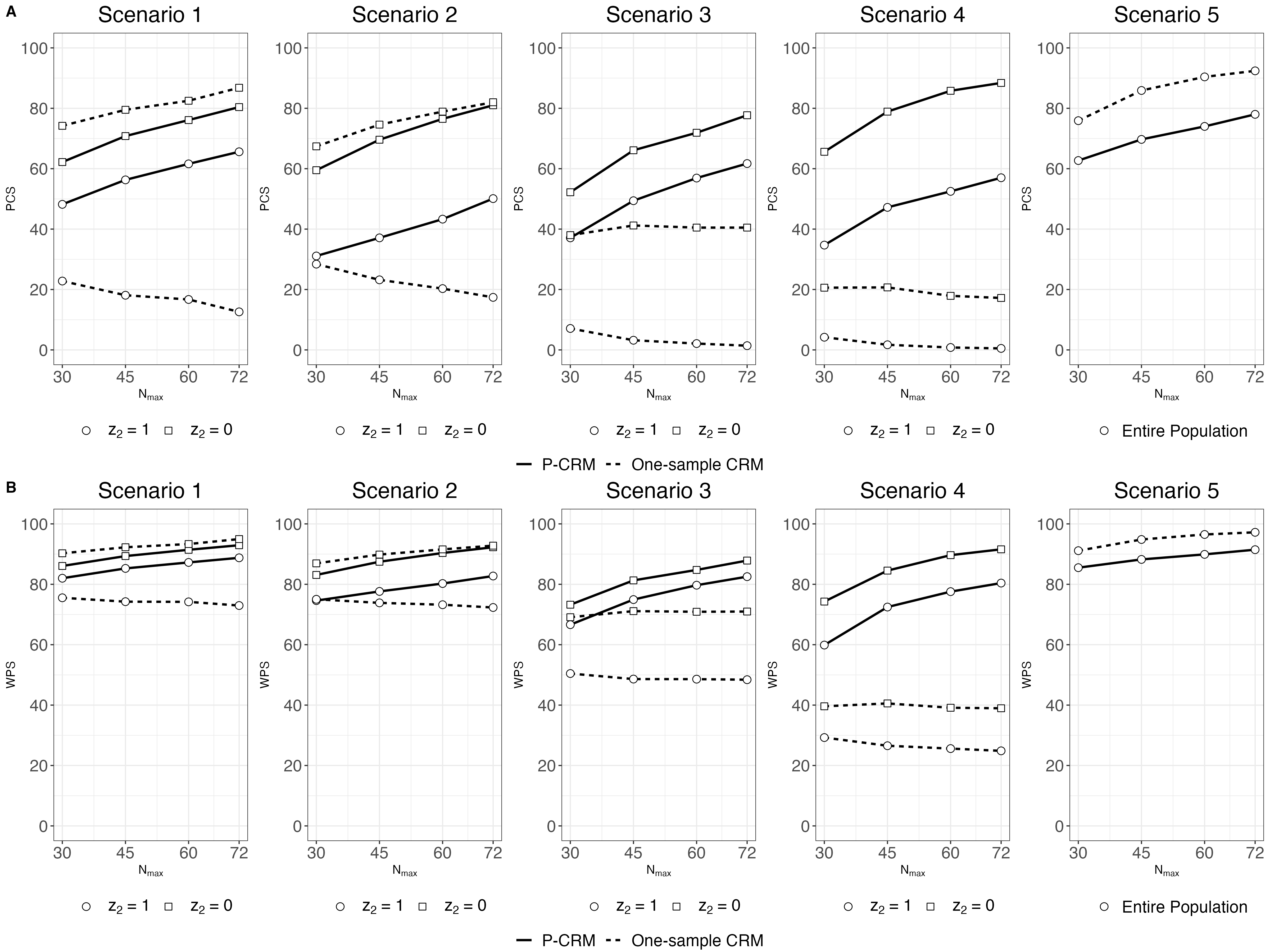}
\caption{A) Proportion of Correct Selection of MTD (PCS) and B) Weighted Probability of Selection (WPS) for each subgroup under the P-CRM and one-sample CRM across $N_{\max} = 30, 45, 60, 72$ for criteria prevalence of 25\%.}
\label{pcs_wps_fig25}
\end{figure}

The probability of criteria selection for the P-CRM when the prevalence of each criterion is 50\% and 25\% is given in Table~\ref{tb:pseltable} for $N_{\max} = 45$.  The PCS and WPS for the P-CRM and one-sample CRM across total sample size for prevalence of each criterion of 50\% and 25\% are shown in Figures~\ref{pcs_wps_fig50} and~\ref{pcs_wps_fig25}, respectively. Additional simulation results for criteria selection for $N_{\max} = 60, 72$ and probability of selection of each dose under the P-CRM and one-sample CRM for a prevalence of 50\% can be found in the Appendix.


Table~\ref{tb:pseltable} gives the probability of criteria selection when $P(z_m = 1) =0.50, 0.25, m = 1,2,3$ and $N_{\max} = 45$. The probability of correct criteria selection is in bold; in Scenarios 1 through 4, only one criterion, $z_2$, should be selected, and in Scenario 5, no criteria should be selected. The P-CRM demonstrates higher criteria selection when the dose-toxicity relationships between subpopulations are more heterogeneous. For example, for a prevalence of 50\%, when the MTDs are at least two dose levels apart, the design has a high selection rate of the true criterion, $z_2$, of around 70\%, but in the more challenging case of Scenarios 1 and 2, where the MTDs are only one dose apart, the method selects $z_2$ between 40 to 50\% of the time. In these cases, when the method does not correctly select the true criterion, it will most often select no criteria, defaulting to the naive method of not including any covariates in the dose-toxicity model and assuming a homogeneous population. In Scenario 5, when no criteria are associated with toxicity, the method selects none of them about 56\% of the time. When the prevalence of each criterion is 25\%, the percentage of correct criterion selection remains similar, slightly decreasing for Scenarios 1 through 3 by an average of 3\%. For larger sample sizes of $N_{\max} = 60$ and $N_{\max} = 72$, the probability of correct criterion selection improves, particularly for the cases of only slight heterogeneity, shown in the Appendix. 

Given that the design has no prior assumptions about which criterion are associated with toxicity, the design selects the correct criterion at a desirable rate for heterogeneous populations where the optimal doses differ by more than one dose. The design has a controlled probability of incorrect criteria selection or false positive rate (FPR), except for Scenario 5 when the population is homogeneous. If the investigator wishes to lower the FPR, they could consider a more conservative adjustment of $\alpha$, but the design will be less likely to screen for patient heterogeneity. When the prevalence decreases by half, the method performs almost as well in selecting criteria as when the criteria prevalence is 50\%, implying the correct selection rate will only fall substantially when the criteria are very rare.

\begin{table}
 \caption{Probability of Criteria Selection with three total covariates, M = 3, \\each with prevalence of 50\% and 25\% for $N_{\max} = 45$}
 \label{tb:pseltable}
  \centering 
     \begin{tabular}{ l  c c c c c}
   \toprule \textbf{Prevalence = 50\%} & Scenario & No criteria & Correct criterion & Correct criterion with others & Incorrect criteria \\  \hline
   & 1 &  30 &  \textbf{48}  &  6  &    16 \\
   & 2 & 30 &    \textbf{44}&    6 &   19  \\  
   & 3  &  11 &  \textbf{68}&    10 &  11  \\
   & 4  &  6 &  \textbf{73}&    14&   7  \\
   & 5 &    \textbf{56} & \small{NA}  & \small{NA}  & 44 \\  \midrule
   
    \textbf{Prevalence = 25\%}  & Scenario & No criteria & Correct criterion & Correct criterion with others & Incorrect criteria \\  \hline
 & 1 &  38 &  \textbf{43}  &  4  &    16 \\
 & 2 & 39 &    \textbf{41}&    3 &   16  \\ 
   & 3 & 15 &    \textbf{67}&    8 &   10  \\  
   & 4  &   7 &  \textbf{79}&    10&   4  \\
   & 5 &   \textbf{63} &  \small{NA} &  \small{NA} &37 \\ 
   \bottomrule
   \end{tabular}
   \begin{tablenotes}
\item[*] Selection probabilities may not equal 100 due to rounding.
\end{tablenotes}
\label{psel_table}
\end{table}


Figure~\ref{pcs_wps_fig50} (A and B) compare the subpopulation-specific PCS and WPS, respectively, between the P-CRM and one-sample CRM when prevalence of each criterion is 50\% across $N_{\max} = 30, 45, 60, 72$. As heterogeneity of the subgroups grows and as total sample size grows, the P-CRM demonstrates substantially improved PCS and WPS over the one-sample CRM. Performance of the one-sample CRM indicates that when there is more than one true MTD, ignoring patient criteria will lead to incorrectly dosing at least one subgroup, and often both. When the subgroups differ in true MTD by at least two doses, the one-sample CRM most often assigns a potentially sub-therapeutic dose to one subgroup and unsafe dose to the other, misidentifying each true respective MTD, whereas the P-CRM most often assigns the true subgroup MTD. Under the P-CRM, the PCS increases more over $N_{\max}$ for more heterogeneous populations. For scenarios where heterogeneity exists, the average PCS exceeds 60\% by $N_{\max} = 60$. For a homogeneous population, the PCS is close to 60\% by $N_{\max} = 30$. The high WPS for all scenarios indicates that the P-CRM rarely assigns a dose that is more than one dose away from the true MTD. Detail of selection rates across each dose for the P-CRM and one-sample CRM are given in Appendix Tables 3 and 4 for criteria prevalence of 50\%, showing that the subgroup with a lower MTD is more likely to be recommended a dose lower than the subgroup without the expanded eligibility criterion when the exact MTD assignment in incorrect.

Figure~\ref{pcs_wps_fig25} (A and B) compare the PCS and WPS, respectively, between the P-CRM and the one-sample CRM when the prevalence of each criterion is 25\% across $N_{\max} = 30, 45, 60, 72$. Compared to when the prevalence is 50\%, the PCS for the subgroups that satisfy the broadened eligibility criteria ($z_i = 1, i = 1,2$) is lower on average by 14\%. However, even though estimating the exact subpopulation-specific MTD for this subgroup is more challenging, the design is more likely to assign a lower dose to the less tolerant subgroup and therefore recommend a dose closer to the true MTD, as evidenced by the still favorable WPS compared to the one-sample CRM. Under the one-sample CRM when the criteria prevalence is 25\%, the subgroup satisfying $z_2 =1$ is completely ignored and for any heterogeneity where the subgroups have MTDs more than one dose apart, the PCS is close to 0\% with a decreasing WPS as heterogeneity increases.


Overall, the P-CRM performs well when there is one criterion differentiating the optimal dose for the patient population when the probability of toxicity at each subgroup’s MTD differs by at least 20\%. The design identifies the true criterion more often for more heterogeneous populations. Regarding dose assignment, we find that even when the design does not correctly identify the true MTD, particularly when the prevalence of the criteria is low, the design tends to recommend a dose that is close to the true MTD rather than assign the same dose as the one recommended for the more prevalent subgroup. Finally, the poor performance of the one-sample CRM highlights how assuming patient homogeneity and ignoring patient criteria leads to overly toxic or sub-therapeutic dose recommendations.

\subsection{Sample Size Considerations}
\label{s:ss}

The total sample size required for the design depends on the number of eligibility criteria that are being considered, $M$, the number of criteria for which we need to differentiate the MTD, $K$, the number of dose levels, $J$, and the prevalence of the criteria. With $J=6$, $M = 3$, a criteria prevalence of 50\%, and only one covariate being associated with toxicity, to achieve an average PCS of at least 50\% across scenarios, the sample size would need to be at least 45. When more than one out of the three covariate is associated with toxicity, resulting in three or more subgroups with different MTDs, a sample size around 60 is needed (data not shown). When prevalence decreases, the overall PCS increases but the PCS of the less prevalent subgroup decreases, and a larger sample size will be needed to achieve a subgroup PCS of at least 50\%. Given the same scenarios, when the number of criteria to consider is only two, $M = 2$, a lower sample size around 35 would obtain an average PCS of at least 50\% (data not shown). If only one criteria is being considered, then other existing methods should also be considered. That is, the particular advantage of this method is that it can consider multiple possible subpopulations at once and thus reduce the required sample size to find one or more MTDs. While learning about heterogeneity requires an increased sample size, conducting separate dose-finding trials for each potential subgroup using the one-sample CRM would require a substantially larger sample size since there are many possible patient subgroups ($2^M$), some of which could have insufficient sample sizes due to low prevalence.


\section{Discussion}
\label{s:discuss}

The precision dose-finding design, P-CRM, proposed in this paper explores dose-finding when the patient population is suspected to be heterogeneous but it is unknown which patient covariates indeed affect toxicity, and thus must be accounted for in MTD determination. As eligibility criteria expand in cancer trials, safety profiles for newly eligible patients could differ given that many were originally included for safety and a method accounting for eligibility criteria is needed to ensure patient safety. One recommendation for how to deal with the broadening of eligibility criteria in early-stage trials is to use expansion cohorts after an initial traditional dose-finding trial [\cite{malikEligibilityCriteriaPhase2019, kimBroadeningEligibilityCriteria2017}]. However, the number of potential cohorts could be large and the specific cohorts that should be included are unknown. The second stage of the P-CRM is comparable to an expansion cohort in that it accounts for potential subpopulations, but the P-CRM is also able to adjust the dose for those populations accurately through a continued dose-finding procedure and learn more about subpopulations that are truly heterogeneous.  

The P-CRM uses marginal logistic regression models to determine whether each covariate is associated with toxicity. Covariates are sequentially added to the dose-toxicity model if they demonstrate significant association with toxicity and are assessed for removal after patient data from each cohort is obtained and other selected covariates are accounted for. Dose assignment and MTD recommendation are based on dose and any patient covariates selected in the model. When no patient covariates are added, the design reverts to the original CRM and assumes patient homogeneity. The design requires no assumption about which covariates differentiate the patient population and can screen for multiple potential covariates at a time. Given our potentially high dimensional model with a sparsity constraint, a natural approach might be the least absolute shrinkage and selection operator (LASSO) method. However, using the LASSO method in the setting of dose-finding where the sample size is small and the data is sparse, led to high false positive rates in covariate selection and an overall smaller probability of correct  MTD selection.


Our simulation study found that the design can correctly identify the covariate associated with toxicity while maintaining control over the false positive rate. Although the design performs most favorably when large patient heterogeneity exists, we find that across any patient heterogeneity scenario, the P-CRM is still favorable over the naive approach of ignoring patient heterogeneity and MTD distinction is necessary. When the probability of correct covariate selection was less favorable due to less heterogeneous subpopulations, the probability of choosing a desirable dose was still high since the true MTDs in that case are close together. Furthermore, the P-CRM did not as accurately choose the true MTD for the less prevalent subgroup, but it chose a dose closer to their respective true MTD than the true MTD of the more prevalent subgroup. Finally, the simulation study showed that ignoring patient characteristics can lead to overly toxic or sub-therapeutic dose recommendations for at least one but possibly all subpopulations within the patient population. Thus, ignoring patient characteristics for a heterogeneous population is far worse than incorporating patient characteristics for a homogeneous population. 

Since our design is the first phase I design to consider expanded eligibility criteria and assume no prior knowledge on heterogeneity, we do not compare it to other methods that account for known patient heterogeneity. For example, the two-sample CRM proposed by \citet{oquigleyTwosampleContinualReassessment1999} or the Hierarchical Bayesian CRM proposed by \citet{moritaSimulationStudyMethods2017} assume that the patient subpopulations are already known. The P-CRM must first identify the subpopulations and then determine the dose for each one.

Although larger sample sizes are needed to accommodate the expansion of many patient criteria compared to traditional phase I trials, the advantage of accounting for heterogeneity even with a typical phase I sample size of 30 is already clear. Furthermore, with the broadening of eligibility, more patients will be available to recruit into the trial and benefit from a safe dose. Thus, it is important to weigh the desired number of criteria to be evaluated and the potential number of subgroups expected against the increase in sample size and the prevalence of the criteria when applying the method in practice. Ongoing work involves including efficacy so that both heterogeneity in toxicity and efficacy can be accounted for to determine the optimal dose.  While precision medicine approaches have been considered mainly for late stage trials and large sample size settings, our approach presents a precision dose-finding approach for early stage trials with a limited sample size.  


The proposed design, P-CRM, selects patient characteristics that are associated with toxicity and determines a subpopulation-specific MTD in a simple manner. The design only requires use of logistic regression, so incorporation of precision medicine in this case does not necessitate specification of multiple priors. A more precise determination of MTD and identification of the source of patient heterogeneity can be an asset in the long term, leading to more successful phase II and III trials and being directly applicable to the true patient population. 

\clearpage
\renewrobustcmd{\bfseries}{\fontseries{b}\selectfont}
\renewrobustcmd{\boldmath}{}
\newrobustcmd{\B}{\bfseries}

\begin{appendix}
\section*{}\label{appn}
\begin{table}[htp]
\centering 
\caption{Probability of toxicity for each dose for Scenarios 1-5.}
  \begin{threeparttable}
\begin{tabular}{llllllll}
\toprule
Scenario &  Subpopulation & D1 & D2 & D3 &D4 & D5 &D6 \\ \midrule

1 & $z_2 = 1$ & \B 0.25 & 0.45 & 0.60 & 0.75 & 0.85 & 0.90\\ 
      & $z_2 = 0$ & 0.02 & \B 0.25 & 0.45 & 0.60 & 0.75 & 0.85 \\ \midrule

2 & $z_2 = 1$ & 0.05 & \B 0.25 &  0.45 & 0.60 & 0.75 & 0.85 \\ 
   & $z_2 = 0$ & 0.02 & 0.05  & \B 0.25 & 0.45 & 0.60 & 0.75\\ \midrule
   
3 & $z_2 = 1$ & 0.05 & \B 0.25 &  0.45 & 0.60 & 0.75 & 0.85 \\ 
   & $z_2 = 0$ & 0.02 & 0.05  & 0.08 & \B 0.25 & 0.45 & 0.60\\ \midrule
      
4 & $z_2 = 1$ & 0.05 & 0.08 & \B 0.25 & 0.45 & 0.60 & 0.70\\ 
       & $z_2 = 0$ & 0.01 & 0.01 & 0.02 & 0.05 & 0.08 & \B 0.25\\ \midrule
      
5 & Entire Population & 0.08 & \B 0.25 & 0.45 & 0.60 & 0.70 & 0.75\\
\bottomrule
\end{tabular}
\end{threeparttable}
\end{table}

\begin{table}[htp]
 \caption{Probability of Criteria Selection with three total covariates, M = 3, \\ each with prevalence of 50\% and 25\% for $N_{\max} = 60$}
  \centering 
    \begin{tabular}{ l  c c c c c}
   \toprule \textbf{Prevalence = 50\%} & Scenario & No criteria & Correct criterion & Correct criterion with others & Incorrect criteria \\  \hline
   & 1 &  21 &  \textbf{57}  &  8  &    14 \\
   & 2  & 24 &    \textbf{52}&    7&   16 \\  
   & 3 & 4 &    \textbf{76}&    14 &   6  \\  
   & 4 &  3 &  \textbf{76}&    19 &   2  \\
   & 5 &    \textbf{56} & \small{NA}  & \small{NA}  & 44 \\  \midrule

    \textbf{Prevalence = 25\%}  & Scenario & No criteria & Correct criterion & Correct criterion with others & Incorrect criteria \\  \hline
 & 1 &  29&  \textbf{50}  &  6  &    16 \\
 & 2  & 31 &    \textbf{48}&    5 &  16  \\  
   & 3 & 8 &    \textbf{73}&    11 &   8  \\  
   & 4  &   2 &  \textbf{82}&    14 &   2  \\
   & 5 &   \textbf{58} &  \small{NA} &  \small{NA} &42 \\  
   \bottomrule
   \end{tabular}
   \begin{tablenotes}
\item[*] Selection probabilities may not equal 100 due to rounding.
\end{tablenotes}
\label{psel_table2}
\end{table}

\begin{table}[htp]
 \caption{Probability of Criteria Selection with three total covariates, M = 3, \\each with prevalence of 50\% and 25\% for $N_{\max} = 72$}
  \centering 
    \begin{tabular}{ l  c c c c c}
   \toprule \textbf{Prevalence = 50\%} & Scenario & No criteria & Correct criterion & Correct criterion with others & Incorrect criteria \\  \hline
   & 1 &  18 &  \textbf{64}  &  8  &    10 \\
   & 2  & 21 &    \textbf{58}&    9 &  13  \\  
   & 3 & 2&    \textbf{78}&    16&   4 \\  
   & 4  &  2 &  \textbf{77}&    20&   1  \\
   & 5 &    \textbf{56} & \small{NA}  & \small{NA}  & 44 \\  \midrule
    
    \textbf{Prevalence = 25\%}  & Scenario & No criteria & Correct criterion & Correct criterion with others & Incorrect criteria \\  \hline
 & 1 &  26 &  \textbf{54}  &  6  &    14 \\
   & 2  & 27 &    \textbf{52}&    6 &  14  \\  
   & 3 & 6&    \textbf{76}&    13 &   5  \\  
   & 4  &   1 &  \textbf{83}&    15&   1  \\
    & 5 &   \textbf{55} &  \small{NA} &  \small{NA} &45 \\  
   \bottomrule
   \end{tabular}
   \begin{tablenotes}
\item[*] Selection probabilities may not equal 100 due to rounding.
\end{tablenotes}
\label{psel_table3}
\end{table}



\begin{table}
\centering
\caption{The proportion of dose selection, PCS, and WPS for each $N_{\max}$ under the P-CRM \\when the criteria prevalence is 50\%.}
\begin{tabular}[t]{llrrrrrrrrrr}
\toprule
P-CRM & Scenario & $N_{\max}$ & True MTD & D1 & D2 & D3 & D4 & D5 & D6 & PCS & WPS\\ \midrule

 & Scenario 1 & 30 & 1 & \B 0.64 & 0.30 & 0.05 & 0.01 & 0.00 & 0.00 & 0.64 & 0.87\\

 &  & 30 & 2 & 0.27 &  \B 0.46 & 0.22 & 0.04 & 0.01 & 0.00 & 0.46 & 0.79\\

 &  & 45 & 1 &  \B 0.71 & 0.25 & 0.03 & 0.00 & 0.00 & 0.00 & 0.71 & 0.90\\

 &  & 45 & 2 & 0.20 &  \B 0.53 & 0.23 & 0.03 & 0.00 & 0.00 & 0.53 & 0.82\\

 &  & 60 & 1 &  \B 0.76 & 0.22 & 0.02 & 0.00 & 0.00 & 0.00 & 0.76 & 0.92\\

 &  & 60 & 2 & 0.16 &  \B 0.62 & 0.21 & 0.02 & 0.00 & 0.00 & 0.62 & 0.86\\

 &  & 72 & 1 &  \B 0.80 & 0.19 & 0.01 & 0.00 & 0.00 & 0.00 & 0.80 & 0.94\\

 &  & 72 & 2 & 0.12 &  \B 0.69 & 0.18 & 0.01 & 0.00 & 0.00 & 0.69 & 0.89\\ \midrule

 & Scenario 2 & 30 & 2 & 0.13 &  \B 0.49 & 0.32 & 0.05 & 0.01 & 0.00 & 0.49 & 0.81\\

 &  & 30 & 3 & 0.03 & 0.28 &  \B 0.44 & 0.20 & 0.05 & 0.00 & 0.44 & 0.76\\

 &  & 45 & 2 & 0.13 &  \B 0.58 & 0.25 & 0.04 & 0.00 & 0.00 & 0.58 & 0.85\\

 &  & 45 & 3 & 0.01 & 0.22 &  \B 0.51 & 0.22 & 0.04 & 0.00 & 0.51 & 0.79\\

 &  & 60 & 2 & 0.10 &  \B 0.65 & 0.22 & 0.02 & 0.00 & 0.00 & 0.65 & 0.88\\

 &  & 60 & 3 & 0.00 & 0.19 &  \B 0.58 & 0.20 & 0.02 & 0.00 & 0.58 & 0.82\\

 &  & 72 & 2 & 0.08 &  \B 0.72 & 0.19 & 0.01 & 0.00 & 0.00 & 0.72 & 0.90\\

 &  & 72 & 3 & 0.00 & 0.17 &  \B 0.64 & 0.17 & 0.01 & 0.00 & 0.64 & 0.85\\ \midrule

 & Scenario 3 & 30 & 2 & 0.10 &  \B 0.49 & 0.31 & 0.08 & 0.02 & 0.00 & 0.49 & 0.76\\

 &  & 30 & 4 & 0.01 & 0.13 & 0.28 &  \B 0.35 & 0.18 & 0.05 & 0.35 & 0.63\\

 &  & 45 & 2 & 0.11 &  \B 0.62 & 0.22 & 0.04 & 0.01 & 0.00 & 0.62 & 0.83\\

 &  & 45 & 4 & 0.00 & 0.06 & 0.20 &  \B 0.48 & 0.21 & 0.04 & 0.48 & 0.70\\

 &  & 60 & 2 & 0.09 &  \B 0.70 & 0.17 & 0.03 & 0.00 & 0.00 & 0.70 & 0.87\\

 &  & 60 & 4 & 0.00 & 0.03 & 0.16 &  \B 0.61 & 0.17 & 0.03 & 0.61 & 0.78\\

 &  & 72 & 2 & 0.07 &  \B 0.76 & 0.15 & 0.01 & 0.00 & 0.00 & 0.76 & 0.90\\

 &  & 72 & 4 & 0.00 & 0.02 & 0.14 &  \B 0.66 & 0.17 & 0.01 & 0.66 & 0.81\\ \midrule

 & Scenario 4 & 30 & 3 & 0.01 & 0.10 &  \B 0.42 & 0.32 & 0.11 & 0.04 & 0.42 & 0.69\\

 &  & 30 & 6 & 0.00 & 0.01 & 0.11 & 0.18 & 0.24 &  \B 0.45 & 0.45 & 0.56\\

 &  & 45 & 3 & 0.00 & 0.12 &  \B 0.55 & 0.26 & 0.05 & 0.01 & 0.55 & 0.79\\

 &  & 45 & 6 & 0.00 & 0.00 & 0.03 & 0.07 & 0.24 &  \B 0.65 & 0.65 & 0.74\\

 &  & 60 & 3 & 0.00 & 0.11 &  \B 0.63 & 0.23 & 0.02 & 0.00 & 0.63 & 0.83\\

 &  & 60 & 6 & 0.00 & 0.00 & 0.01 & 0.04 & 0.21 &  \B 0.74 & 0.74 & 0.81\\

 &  & 72 & 3 & 0.00 & 0.10 &  \B 0.68 & 0.20 & 0.01 & 0.00 & 0.68 & 0.86\\

 &  & 72 & 6 & 0.00 & 0.00 & 0.01 & 0.03 & 0.19 &  \B 0.78 & 0.78 & 0.84\\ \midrule

 & Scenario 5 & 30 & 2 & 0.20 & \B  0.57 & 0.19 & 0.04 & 0.00 & 0.00 & 0.57 & 0.83\\

 &  & 45 & 2 & 0.17 &  \B 0.63 & 0.17 & 0.02 & 0.00 & 0.00 & 0.63 & 0.85\\

 &  & 60 & 2 & 0.15 &  \B 0.69 & 0.15 & 0.01 & 0.00 & 0.00 & 0.69 & 0.88\\

 &  & 72 & 2 & 0.12 & \B  0.73 & 0.14 & 0.01 & 0.00 & 0.00 & 0.73 & 0.90\\
\bottomrule
\end{tabular}
\end{table}

\clearpage

\begin{table}
\centering
\caption{The proportion of dose selection, PCS, and WPS for each $N_{\max}$ under the one-sample CRM \\when the criteria prevalence is 50\%}
\begin{tabular}[t]{llrrrrrrrrrr}
\toprule
One-sample CRM & Scenario & $N_{\max}$ & True MTD & D1 & D2 & D3 & D4 & D5 & D6 & PCS & WPS\\ \toprule

 & Scenario 1 & 30 & 1 & \B 0.46 & 0.52 & 0.02 & 0.00 & 0.00 & 0.00 & 0.46 & 0.83\\

 &  & 30 & 2 & 0.46 & \B 0.52 & 0.02 & 0.00 & 0.00 & 0.00 & 0.52 & 0.82\\

 &  & 45 & 1 & \B 0.46 & 0.54 & 0.00 & 0.00 & 0.00 & 0.00 & 0.46 & 0.83\\

 &  & 45 & 2 & 0.46 & \B 0.54 & 0.00 & 0.00 & 0.00 & 0.00 & 0.54 & 0.82\\

 &  & 60 & 1 & \B 0.48 & 0.52 & 0.00 & 0.00 & 0.00 & 0.00 & 0.48 & 0.84\\

 &  & 60 & 2 & 0.48 & \B 0.52 & 0.00 & 0.00 & 0.00 & 0.00 & 0.52 & 0.82\\

 &  & 72 & 1 & \B 0.48 & 0.52 & 0.00 & 0.00 & 0.00 & 0.00 & 0.48 & 0.84\\

 &  & 72 & 2 & 0.48 & \B 0.52 & 0.00 & 0.00 & 0.00 & 0.00 & 0.52 & 0.82\\ \midrule

 & Scenario 2 & 30 & 2 & 0.00 & \B 0.52 & 0.46 & 0.02 & 0.00 & 0.00 & 0.52 & 0.84\\

 &  & 30 & 3 & 0.00 & 0.52 &\B  0.46 & 0.02 & 0.00 & 0.00 & 0.46 & 0.78\\

 &  & 45 & 2 & 0.00 & \B 0.52 & 0.48 & 0.00 & 0.00 & 0.00 & 0.52 & 0.84\\

 &  & 45 & 3 & 0.00 & 0.52 & \B 0.48 & 0.00 & 0.00 & 0.00 & 0.48 & 0.79\\

 &  & 60 & 2 & 0.00 & \B 0.52 & 0.47 & 0.00 & 0.00 & 0.00 & 0.52 & 0.84\\

 &  & 60 & 3 & 0.00 & 0.52 & \B 0.47 & 0.00 & 0.00 & 0.00 & 0.47 & 0.79\\

 &  & 72 & 2 & 0.00 & \B 0.52 & 0.47 & 0.00 & 0.00 & 0.00 & 0.52 & 0.84\\

 &  & 72 & 3 & 0.00 & 0.52 & \B 0.47 & 0.00 & 0.00 & 0.00 & 0.47 & 0.79\\ \midrule

 & Scenario 3 & 30 & 2 & 0.00 & \B 0.30 & 0.56 & 0.14 & 0.00 & 0.00 & 0.30 & 0.68\\

 &  & 30 & 4 & 0.00 & 0.30 & 0.56 & \B 0.14 & 0.00 & 0.00 & 0.14 & 0.55\\

 &  & 45 & 2 & 0.00 & \B 0.25 & 0.65 & 0.10 & 0.00 & 0.00 & 0.25 & 0.67\\

 &  & 45 & 4 & 0.00 & 0.25 & 0.65 & \B 0.10 & 0.00 & 0.00 & 0.10 & 0.54\\

 &  & 60 & 2 & 0.00 & \B 0.23 & 0.70 & 0.07 & 0.00 & 0.00 & 0.23 & 0.67\\

 &  & 60 & 4 & 0.00 & 0.23 & 0.70 & \B 0.07 & 0.00 & 0.00 & 0.07 & 0.53\\

 &  & 72 & 2 & 0.00 & \B 0.22 & 0.72 & 0.05 & 0.00 & 0.00 & 0.22 & 0.67\\

 &  & 72 & 4 & 0.00 & 0.22 & 0.72 & \B 0.05 & 0.00 & 0.00 & 0.05 & 0.52\\ \midrule

 & Scenario 4 & 30 & 3 & 0.00 & 0.01 & \B 0.24 & 0.52 & 0.21 & 0.03 & 0.24 & 0.58\\

 &  & 30 & 6 & 0.00 & 0.01 & 0.24 & 0.52 & 0.21 & \B 0.03 & 0.03 & 0.19\\

 &  & 45 & 3 & 0.00 & 0.00 & \B 0.19 & 0.57 & 0.22 & 0.01 & 0.19 & 0.56\\

 &  & 45 & 6 & 0.00 & 0.00 & 0.19 & 0.57 & 0.22 & \B 0.01 & 0.01 & 0.18\\

 &  & 60 & 3 & 0.00 & 0.00 & \B 0.18 & 0.61 & 0.20 & 0.01 & 0.18 & 0.56\\

 &  & 60 & 6 & 0.00 & 0.00 & 0.18 & 0.61 & 0.20 & \B 0.01 & 0.01 & 0.17\\

 &  & 72 & 3 & 0.00 & 0.00 & \B 0.15 & 0.66 & 0.19 & 0.00 & 0.15 & 0.55\\

 &  & 72 & 6 & 0.00 & 0.00 & 0.15 & 0.66 & 0.19 & \B 0.00 & 0.00 & 0.18\\ \midrule

 & Scenario 5 & 30 & 2 & 0.13 & \B 0.76 & 0.11 & 0.00 & 0.00 & 0.00 & 0.76 & 0.91\\

 &  & 45 & 2 & 0.08 & \B 0.86 & 0.06 & 0.00 & 0.00 & 0.00 & 0.86 & 0.95\\

 &  & 60 & 2 & 0.06 & \B 0.90 & 0.04 & 0.00 & 0.00 & 0.00 & 0.90 & 0.96\\

 &  & 72 & 2 & 0.04 & \B 0.92 & 0.03 & 0.00 & 0.00 & 0.00 & 0.92 & 0.97\\
\bottomrule
\end{tabular}
\end{table}

\end{appendix}
%
%


\clearpage

\begin{acks}[Acknowledgments]
This publication was supported by the National Center for Advancing Translational Sciences, National Institutes of Health, through Grant Numbers UL1TR001873 and TL1TR001875.
\end{acks}


\begin{thebibliography}{4}
\bibitem[\protect\citeauthoryear{Babb and Rogatko}{2001}]{babbPatientSpecificDosing2001} \textsc{Babb, J. S. and Rogatko, A.} (2001). Patient specific dosing in a cancer phase I clinical trial. {\it Statistics in Medicine} {\bf 20,} 2079–2090. 
\bibitem[\protect\citeauthoryear{Beaver, Ison, and Pazdur}{Beaver et al.}{2017}]{beaverReevaluatingEligibilityCriteria2017a} \textsc{Beaver, J. A., Ison, G., and Pazdur, R.} (2017). Reevaluating Eligibility Criteria — Balancing Patient Protection and Participation in Oncology Trials. {\it New England Journal of Medicine} {\bf 16,} 1504–1505. 
\bibitem[\protect\citeauthoryear{Chapple and Thall}{2018}]{chappleSubgroupspecificDoseFinding2018} \textsc{Chapple, A. D. and Thall, P. F.} (2018). Subgroup-specific dose finding in phase I clinical trials based on time to toxicity allowing adaptive subgroup combination. {\it Pharmaceutical statistics} {\bf 17,} 734–749. 
\bibitem[\protect\citeauthoryear{Cheng et al.}{2004}]{chengIndividualizedPatientDosing2004} \textsc{Cheng, J. D., Babb, J. S., Langer, C., Aamdal, S., Robert, F., Engelhardt, L.R., et al.} (2004). Individualized patient dosing in phase I clinical trials: the role of escalation with overdose control in PNU-214936. {\it Journal of Clinical Oncology: Official Journal of the American Society of Clinical Oncology} {\bf 22,} 602–609. 


\bibitem[\protect\citeauthoryear{Conaway}{2017}]{conawayIsotonicDesignsPhase2017} \textsc{Conaway, M.} (2017). Isotonic designs for phase I trials in partially ordered groups. {\it Clinical Trials} {\bf 14,} 491–498.

\bibitem[\protect\citeauthoryear{Fehrenbacker, Ackerson, and Somkin}{Fehrenbacker et al.}{2009}]{fehrenbacherRandomizedClinicalTrial2009} \textsc{Fehrenbacher, L., Ackerson, L., and Somkin, C.} (2009). Randomized clinical trial eligibility rates for chemotherapy (CT) and antiangiogenic therapy (AAT) in a population-based cohort of newly diagnosed non-small cell lung cancer (NSCLC) patients. {\it Journal of Clinical Oncology} {\bf 27,} 6538.  
\bibitem[\protect\citeauthoryear{George}{1996}]{georgeReducingPatientEligibility1996} \textsc{George, S. L.} (1996). Reducing patient eligibility criteria in cancer clinical trials. {\it Journal of Clinical Oncology: Official Journal of the American Society of Clinical Oncology} {\bf 14, }1364–1370. 
\bibitem[\protect\citeauthoryear{Gerber et al.}{2016}]{gerberImpactPriorCancer2014} \textsc{Gerber, D. E., Laccetti, A.L., Xuan, L., Halm, E. A., and Pruitt, S. L.} (2016). Impact of Prior Cancer on Eligibility for Lung Cancer Clinical Trials. {\it Journal of the National Cancer Institute} {\bf 106,} 302.
\bibitem[\protect\citeauthoryear{Harvey et al.}{2021}]{harveyImpactBroadeningTrial2021} \textsc{Harvey, R. D., Mileham, K. F., Bhatnagar, V., Brewer, J. R., Rahman, A., Moravek, C., et al.} (2021). Impact of Broadening Trial Eligibility Criteria for Patients with Advanced Non-Small Cell Lung Cancer: Real-World Analysis of Select ASCO-Friends Recommendations. {\it Clinical cancer research} {\bf 27,} 2430–2434. 
\bibitem[\protect\citeauthoryear{Khan et al.}{2022}]{khanIntermittentMEKInhibition2022} \textsc{Khan, S., Patel, S. P., Shoushtari, A. N., Ambrosini, G., Cremers, S., Lee, S. M., et al.} (2022) Intermittent {MEK} inhibition for the treatment of metastatic uveal melanoma. {\it Frontiers in Oncology } {\bf 12}. doi: 10.3389/fonc.2022.975643.
\bibitem[\protect\citeauthoryear{Kim et al.}{2015}]{kimModernizingEligibilityCriteria2015} \textsc{Kim, E. S., Bernstein, D., Hilsenbeck, S. G., Chung, C. H., Dicker, A. P., Ersek, J. L., et al.}(2015). Modernizing Eligibility Criteria for Molecularly Driven Trials. {\it Journal of Clinical Oncology: Official Journal of the American Society of Clinical Oncology} {\bf 33,} 2815–2820. 
\bibitem[\protect\citeauthoryear{Kim et al.}{2017}]{kimBroadeningEligibilityCriteria2017} \textsc{Kim, E. S., Bruinooge, S. S., Roberts, S., Ison, G., Lin, N. U., Gore, L., et al.} (2017). Broadening Eligibility Criteria to Make Clinical Trials More Representative: American Society of Clinical Oncology and Friends of Cancer Research Joint Research Statement. {\it Journal of Clinical Oncology} {\bf 35,} 3737. 
\bibitem[\protect\citeauthoryear{Kitchlu et al.}{2018}]{kitchluRepresentationPatientsChronic2018} \textsc{Kitchlu, A., Shapiro, J., Amir, E., Garg, A. X., Kim, S. J., et al.} (2018). Representation of Patient with Chronic Kidney Disease. {\it Journal of the American Medical Association} {\bf 319}, 2437-2439. 
\bibitem[\protect\citeauthoryear{Laccetti et al.}{2016}]{laccettiPriorCancerDoes2016} \textsc{Laccetti, A. L., Pruitt, S. L., Xuan, L., Halm, E. A., and Gerber, D. E.} (2016). Prior cancer does not adversely affect survival in locally advanced lung cancer: A national SEER-Medicare analysis. {\it Lung cancer (Amsterdam, Netherlands)} {\bf 98,} 106.  
\bibitem[\protect\citeauthoryear{Lameire}{2014}]{lameireNephrotoxicityRecentAnticancer2014} \textsc{Lameire, N.} (2014). Nephrotoxicity of recent anti-cancer agents. {\it Clinical Kidney Journal} {\bf 7,} 11–22. 
\bibitem[\protect\citeauthoryear{Leal et al.}{2011}]{lealDoseescalatingPharmacologicalStudy2011} \textsc{Leal, T. B., Remick, S. C., Takimoto, C. H., Ramanathan, R. K., Davies, A., Egorin, M. J., et al.} (2011). Dose-escalating and pharmacological study of bortezomib in adult cancer patients with impaired renal function: a National Cancer Institute Organ Dysfunction Working Group Study. {\it Cancer Chemotherapy and Pharmacology} {\bf 68,} 1439–1447. 
\bibitem[\protect\citeauthoryear{Lee and Cheung}{2009}]{leeModelCalibrationContinual2009} \textsc{Lee, S. M. and Cheung, Y. K.} (2009). Model Calibration in the Continual Reassessment Method.  {\it Clinical trials (London, England)} {\bf 6,} 227–238. 
\bibitem[\protect\citeauthoryear{Liu et al.}{2021}]{liuEvaluatingEligibilityCriteria2021} \textsc{Liu, R., Rizzo, S., Whipple, S., Pal, N., Pineda, A. L., Lu, M., et al.} (2021). Evaluating eligibility criteria of oncology trials using real-world data and AI. {\it Nature} {\bf 592,} 629–633. 
\bibitem[\protect\citeauthoryear{LoRusso et al.}{2012}]{lorussoPharmacokineticsSafetyBortezomib2012} \textsc{LoRusso, P. M., Venkatakrishnan, K., Ramanathan, R. K., Sarantopoulos, J., Mulkerin, D., Shibata, S. I., et al.} (2012). Pharmacokinetics and safety of bortezomib in patients with advanced malignancies and varying degrees of liver dysfunction: phase I NCI Organ Dysfunction Working Group Study NCI-6432. {\it Clinical Cancer Research: An Official Journal of the American Association for Cancer Research} {\bf 18,} 2954–2963. 
\bibitem[\protect\citeauthoryear{Magnuson et al.}{2021}]{magnusonModernizingClinicalTrial2021} \textsc{Magnuson, A., Bruinooge, S. S., Singh, H., Wilner, K. D., Jalal, S., Lichtman, S. M., et al.} (2021). Modernizing Clinical Trial Eligibility Criteria: Recommendations of the ASCO-Friends of Cancer Research Performance Status Work Group. {\it Clinical Cancer Research: An Official Journal of the American Association for Cancer Research} {\bf 27,} 2424–2429. 
\bibitem[\protect\citeauthoryear{Malik and Lu}{2019}]{malikEligibilityCriteriaPhase2019} \textsc{Malik, L. and Lu, D.} (2019). Eligibility criteria for phase I clinical trials: tight vs loose? {\it Cancer Chemotherapy and Pharmacology} {\bf 83,} 999–1002. 
\bibitem[\protect\citeauthoryear{Morita, Thall, and Takeda}{Morita et al.}{2017}]{moritaSimulationStudyMethods2017} \textsc{Morita, S., Thall, P. F., and Takeda, K.} (2017). A Simulation Study of Methods for Selecting Subgroup-Specific Doses in Phase I Trials.  {\it Pharmaceutical statistics} {\bf 16,} 143–156. 
\bibitem[\protect\citeauthoryear{O’Quigley and Paoletti}{2003}]{oquigleyContinualReassessmentMethod2003} \textsc{O’Quigley, J. and Paoletti, X.} (2003). Continual Reassessment Method for Ordered Groups. {\it Biometrics} {\bf 59,} 430–440. 
\bibitem[\protect\citeauthoryear{O’Quigley, Pepe, and Fisher}{1990}]{oquigleyContinualReassessmentMethod1990} \textsc{O’Quigley, J., Pepe, M., and Fisher, L.} (1990). Continual Reassessment Method: A practical design for phase I clinical trials in cancer. {\it Biometrics} {\bf 46,} 33–48. 
\bibitem[\protect\citeauthoryear{O’Quigley, Shen, and Gamst}{O’Quigley et al.}{1999}]{oquigleyTwosampleContinualReassessment1999} \textsc{O'Quigley, J., Shen, L. Z. and Gamst, A.} (1999). Two-sample continual reassessment method. {\it  Journal of Biopharmaceutical Statistics} {\bf 9,} 17–44. 
\bibitem[\protect\citeauthoryear{Osarogiagbon  et al.}{2021}]{osarogiagbonModernizingClinicalTrial2021} \textsc{Osarogiagbon, R. U., Vega, D. M., Fashoyin-Aje, L., Wedam, S., Ison, G., Atienza, S., et al.} (2021). Modernizing Clinical Trial Eligibility Criteria: Recommendations of the ASCO-Friends of Cancer Research Prior Therapies Work Group. {\it Clinical Cancer Research: An Official Journal of the American Association for Cancer Research}  {\bf 27,} 2408–2415. 
\bibitem[\protect\citeauthoryear{Piantadosi and Liu}{1996}]{[piantidosi1996]} \textsc{Piantadosi, S. and Liu, G.} (1996). Improved designs for dose escalation studies using pharmacokinetic measurements. {\it Statistics in Medicine} {\bf 15,} 1605–1618. 


\bibitem[\protect\citeauthoryear{Food and Drug Administration}{2020}]{fda} \textsc{U.S. Food and Drug Administration.} (2020). Enhancing the Diversity of Clinical Trial Populations — Eligibility Criteria, Enrollment Practices, and Trial Designs Guidance for Industry. https://www.fda.gov/regulatory-information/search-fda-guidance-documents/enhancing-diversity-clinical-trial-populations-eligibility-criteria-enrollment-practices-and-trial. 

\bibitem[\protect\citeauthoryear{Yuan and Chappell}{2004}]{yuan_chappell2004} \textsc{Yuan, Z. and Chappell, R.} (2004). Isotonic designs for phase I cancer clinical trials with multiple risk groups. {\it Clinical Trials (London, England)} {\bf 1,} 499–508. 



\end{thebibliography}
\end{document}